\documentclass[a4paper,12pt]{article}
\usepackage[cp1251]{inputenc}
\usepackage{amsmath}
\usepackage{amssymb}
\usepackage{textcomp}
\usepackage{euscript}
\usepackage{graphicx}
\usepackage[pdftex]{color}
\usepackage[square,numbers,comma]{natbib}
\usepackage[pdftex]{hyperref}
\hypersetup{
             final,
             colorlinks=true,
             linkcolor=blue,
             citecolor=red
}

\begin{document}

{\bf\textit{Dedicate to L. D. Landau centenary}}

\vspace{0.5cm}

\hspace{7.8cm}{\bf Bulletin of the Lebedev Physics Institute}
\begin{center}
{\bf PLANAR HETEROSTRUCTURE \\ GRAPHENE --- NARROW-GAP SEMICONDUCTOR
--- GRAPHENE}
\end{center}

\vspace{0.3cm}

\hspace{3.5cm}P. V. Ratnikov\footnote{ratnikov@lpi.ru} and A. P.
Silin

\begin{list}{}
{\rightmargin=0cm \leftmargin=4cm}\item\textit{We investigate a
planar heterostructure composed of two graphene films separated by a
narrow-gap semiconductor ribbon. We show that there is no the Klein
paradox when the Dirac points of the Brillouin zone of graphene are
in a band gap of a narrow-gap semiconductor. There is the energy
range depending on an angle of incidence, in which the above-barrier
damped solution exists. Therefore, this heterostructure is a
``filter'' transmitting particles in a certain range of angles of
incidence upon a potential barrier. We discuss the possibility of
application of this heterostructure as a ``switch''.}
\end{list}
\vspace{0.2cm}

Graphene is a two-dimensional gapless semiconductor, and charge
carriers are massless Dirac fermions \cite{Novoselov}. It is known
\cite{Schweber} that a massless relativistic particle with spin 1/2
possesses the chirality property, i.e. it is characterized by a
certain spin projection onto its momentum. In case of graphene, the
chirality is defined by a projection of pseudospin onto a momentum
direction, which is positive for electrons but negative for holes
near K point of the Brillouin zone (BZ) \cite{Katsnelson}, i.e. an
electron and a hole are analogues of a massless neutrino with the
right- and left-hand helicities, respectively. However, the
situation is inverse near K$^\prime$ point where electrons and holes
have the left- and right-hand helicities, respectively,
\cite{McEuen, Neto}. The massless relativistic particle is described
by one spinor, i.e.  a two-component wave function \cite{Landau,
Salam}. It gives basis to state that the effective Hamiltonian
describing the charge carriers in graphene near K point is a
$2\times2$ matrix, and corresponding equation is the Weyl
equation\footnote{The massless fermions considered separately near K
and K$^\prime$ points are similar to the Weyl (two-component)
neutrino. The Dirac equation is used for its simultaneous
description \cite{Ando}. The Dirac equation is equivalent to a pair
of the Weyl equations. The Dirac equation in two-dimensions can be
written as a $2\times2$ matrix (it can be used equivalently with a
$4\times4$ matrix representation) which coincides with the Weyl
equation for a massless particle on a plane. However, the former
equation can be also used for a description of a particle with finite mass.
Using this fact, the problem very close to considered in this paper
task was earlier solved by Gomes and Peres \cite{Peres}.}
\begin{equation}\label{1}
u\boldsymbol{\sigma}\cdot\widehat{\bf p}\,\psi=E\psi,
\end{equation}
where $u=9.84\times10^7$ cm/s is the Fermi velocity which is an
analogue of the Kane matrix element for the rate of interband
transitions in the Dirac model \cite{Ratnikov1}, $\widehat{\bf
p}=-i\hbar\boldsymbol{\nabla}$ (hereafter $\hbar=1$), and
$\boldsymbol{\sigma}=(\sigma_x, \sigma_y)$ are the Pauli matrices.
The dispersion relation of the charge carriers is linear in momentum
$k$
\begin{equation}
E=\pm u k.
\end{equation}

A narrow-gap semiconductor is described by the $4\times4$ matrix
Dirac equation \cite{Volkov}
\begin{equation}\label{3}
\widehat{H}_D\Psi=\left\{\overline{u}\boldsymbol{\alpha}\cdot\widehat{\bf
p}+\beta\Delta+V_0\right\}\Psi=E\Psi,
\end{equation}
where $\Psi$ is a bispinor, $\overline{u}$ is the Kane matrix
element for the rate of interband transitions,
$\boldsymbol{\alpha}=\begin{pmatrix}0& \boldsymbol{\sigma}
\\ \boldsymbol{\sigma}&0\end{pmatrix}$ are the Dirac $\alpha$-matrices, $\beta=\begin{pmatrix} I & 0\\0 & -I
\end{pmatrix}$, $I$ is the $2\times2$ unit matrix, $\Delta$ is half the band gap,
and $V_0$ is the difference of work functions of the narrow-gap
semiconductor and graphene ($|V_0|<\Delta$).

It is necessary to introduce the four-component wave function,
bispinor, for simultaneous description of the charge carriers in
graphene and the narrow-gap semiconductor. In this case, the Dirac
Hamiltonian is
\begin{equation}\label{4}
\widehat{H}_D=\begin{pmatrix}0&u\boldsymbol{\sigma}\cdot\widehat{\bf
p}
\\ u\boldsymbol{\sigma}\cdot\widehat{\bf
p}&0\end{pmatrix}.
\end{equation}
Hamiltonian \eqref{4} is equivalent to Hamiltonian used in Ref.
\cite{Ando}, with an accuracy of two consecutively performed unitary
transformations
$\widehat{U}_2=\frac{1}{\sqrt{2}}\bigl(\begin{smallmatrix} I & I\\I
& -I
\end{smallmatrix}\bigr)$ and $\widehat{U}_1=\bigl(\begin{smallmatrix} I & 0\\0 &
\sigma_y
\end{smallmatrix}\bigr)$ \cite{Ratnikov2}. In addition, both
left-hand and right-hand helicity massless fermions are presented in
the system. Transitions between K and K$^\prime$ points are
improbable, so it is possible to consider that particles conserve
the chirality property.

By performing the unitary transformation $\widehat{U}_2$, it is
convenient to present the Dirac Hamiltonian describing the charge
carriers in the total heterostructure (\hyperlink{fig1}{Fig. 1(a)}) in the form in
which diagonal blocks contain momentum operators
\begin{equation}
\widehat{H}^{\prime}_D=
\begin{pmatrix}
u_i\boldsymbol{\sigma}\cdot\widehat{\bf p}+V_i&\Delta_i\\
\Delta_i&-u_i\boldsymbol{\sigma}\cdot\widehat{\bf p}+V_i
\end{pmatrix},
\end{equation}
where $u_1=u_3=u,\ V_1=V_3=0$, and $\Delta_1=\Delta_3=0$ are the
parameters related to graphene, $u_2=\overline{u},\ V_2=V_0$, and
$\Delta_2=\Delta$ are the parameters of the narrow-gap semiconductor
(\hyperlink{fig1}{Fig. 1(b)}).

For the components of the bispinor describing a particle in
graphene, the following equalities exist
\begin{equation}\label{8}
\begin{split}
\psi_2=s\psi_1e^{i\phi},{}\\
\psi_4=-s\psi_3e^{i\phi},
\end{split}
\end{equation}
where $\phi=\arctan\frac{k_y}{k_x}$ is the polar angle of momentum
${\bf k}=(k_x, k_y)$ of the charge carriers in graphene (the angle
of incidence), $s=signE$.

For the components of the bispinor describing a particle in the
narrow-gap semiconductor, the following equalities exist
\begin{equation}\label{9}
\begin{split}
\psi_3=\frac{E-V_0}{\Delta}\psi_1-\frac{\overline{u}q_x-i\overline{u}k_y}{\Delta}\psi_2,{}\\
\psi_4=-\frac{\overline{u}q_x+i\overline{u}k_y}{\Delta}\psi_1+\frac{E-V_0}{\Delta}\psi_2,
\end{split}
\end{equation}

\hspace{-0.6cm}where
\begin{equation}
\overline{u}^2q^2_x=(E-V_0)^2-\Delta^2-\overline{u}^2k^2_y.
\end{equation}

We find the solution in three ranges: I)\ $x<0$, II)\ $0<x<D$, III)\
$x>D$ ($D$ is the width of the narrow-gap semiconductor ribbon, see
\hyperlink{fig1}{Fig. 1(a)}), taking into account relations \eqref{8}, \eqref{9} and
assuming that the solution is oscillating in range II ($q^2_x>0$),
\begin{equation}
\Psi_I=\begin{pmatrix}c_1\\ sc_1e^{i\phi}\\
c_2\\
-sc_2e^{i\phi}\end{pmatrix}e^{i(k_xx+k_yy)}+\begin{pmatrix}rc_1\\
-s rc_1e^{-i\phi}\\ rc_2\\ s
rc_2e^{-i\phi}\end{pmatrix}e^{i(-k_xx+k_yy)},
\end{equation}
\begin{equation}
\Psi_{II}=\begin{pmatrix}a_1\\ a_2\\
a_1\frac{E-V_0}{\Delta}-a_2\frac{\overline{u}q_x-i\overline{u}k_y}{\Delta}\\
-a_1\frac{\overline{u}q_x+i\overline{u}k_y}{\Delta}+a_2\frac{E-V_0}{\Delta}\end{pmatrix}e^{i(q_xx+k_yy)}+
\begin{pmatrix}
b_1\\ b_2\\
b_1\frac{E-V_0}{\Delta}+b_2\frac{\overline{u}q_x+i\overline{u}k_y}{\Delta}\\
b_1\frac{\overline{u}q_x-i\overline{u}k_y}{\Delta}+b_2\frac{E-V_0}{\Delta}
\end{pmatrix}e^{i(-q_xx+k_yy)},
\end{equation}
\begin{equation}
\Psi_{III}=\begin{pmatrix}tc_1\\ stc_1e^{i\phi}\\
tc_2\\
-stc_2e^{i\phi}\end{pmatrix}e^{i(k_xx+k_yy)},
\end{equation}

\begin{center}
\hypertarget{fig1}{}
\includegraphics[width=10cm]{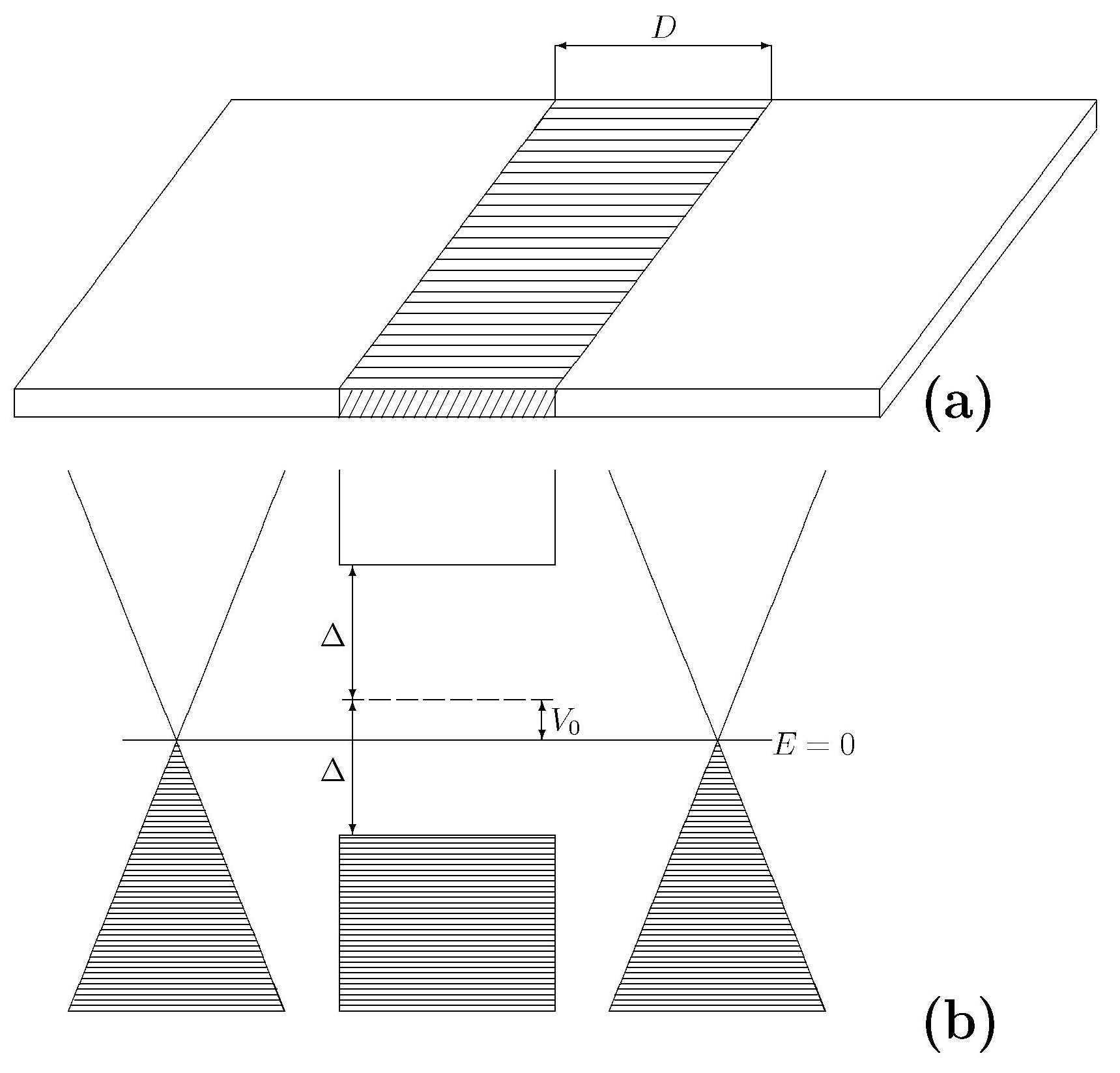}
\end{center}
Fig. 1. \emph{Considered planar heterostructure; (a) two graphene
layers separated by the narrow-gap semiconductor ribbon with the
width D (it is hatched), (b) the band structure: level $E=0$
corresponds to the position of the Dirac points in the BZ of
graphene, the band gap of the narrow-gap semiconductor is
$E_g=2\Delta$, $V_0$ is the difference of the work functions of
graphene and the narrow-gap semiconductor, completely filled valence
bands are hatched.}

\newpage
\hspace{-0.6cm}where $r$ and $t$ are the reflection and transmission
coefficients, respectively, \cite{Katsnelson}, $c_1,$ $c_2,$ $a_1,$
$a_2,\ b_1,\ b_2$ are complex constants determined from the boundary
conditions\footnote{It should be emphasized that $c_2=0$ in the
bispinors $\Psi_I$ and $\Psi_{III}$ for the right-hand helicity
particle, since, the equality $\frac{1-i\gamma_5}{2}\Psi_R=\Psi_R$,
where $\gamma_5=i\beta$, is valid for the bispinor $\Psi_R$
describing the right-hand helicity particle, and $c_1=0$ and the
equality $\frac{1+i\gamma_5}{2}\Psi_L=\Psi_L$ is valid in those
bispinors for the left-hand helicity particle \cite{Schweber}.
Consequently, corresponding components of $\Psi_{II}$ are zero on
interfaces (ones are zero everywhere for the damped solution).}.

Using the boundary conditions \cite{Silin1, Kolesnikov}
\begin{equation}
\sqrt{u^{(-)}}\Psi^{(-)}=\sqrt{u^{(+)}}\Psi^{(+)},
\end{equation}
where the quantities marked by ``($-$)'' and ``($+$)'' relate to
material placing on the left and right of the boundary,
respectively, we obtain for the transmission coefficient
\begin{equation}\label{17}
t=\frac{\cos\phi}{\cos\phi\cos(q_xD)+i\left(\tan\theta\sin\phi-s\frac{E-V_0}{\overline{u}q_x}\right)\sin(q_xD)}e^{-ik_xD},
\end{equation}
where $\tan\theta=\frac{k_y}{q_x}$. Expression \eqref{17}
corresponds to the oscillating solution in range II. In order to get
the transmission coefficient at the exponentially damped solution in
range II, the replacement $q_x\rightarrow i\widetilde{q}_x$ should
be made where
$\overline{u}^2\widetilde{q}^2_x=\Delta^2+\overline{u}^2k^2_y-(E-V_0)^2$,
and $\widetilde{q}^2_x>0$. The transmission probabilities $T=|t|^2$
for both kinds of solution in range II are
\begin{equation}\label{18}
T_\text{oscil}=\frac{\cos^2\phi}{\cos^2\phi\cos^2(q_xD)+\left(\tan\theta\sin\phi-s\frac{E-V_0}
{\overline{u}q_x}\right)^2\sin^2(q_xD)},
\end{equation}
\begin{equation}\label{19}
T_\text{damp}=\frac{\cos^2\phi}{\cos^2\phi\
\cosh^2(\widetilde{q}_xD)+\left(\frac{k_y}{\widetilde{q}_x}\sin\phi-s\frac{E-V_0}
{\overline{u}\widetilde{q}_x}\right)^2\sinh^2(\widetilde{q}_xD)}.
\end{equation}
One can see from formula \eqref{18} that $T_\text{oscil}=1$ when
$q_xD=\pi N$, where $N$ is integer. It corresponds to maxima of the
transmission probability shown in \hyperlink{fig2}{Fig. 2 (a)-(d)}.

As one should expect, the transmission probability in case of the
damped solution in range II is exponentially small for sufficiently
large width of the narrow-gap semiconductor ribbon
$D\gg1/|\widetilde{q}_x|$: $T_\text{damp}\sim
e^{-2|\widetilde{q}_x|D}$. The result of the passage to limit,
$\Delta\rightarrow0$, in \eqref{17} coincides with the transmission
coefficient $t$ in Ref. \cite{Katsnelson}.

The reflection coefficient is simply obtained
\begin{equation}\label{21}
r=-i\sin(q_xD)\frac{\cos(\phi-\theta)-s\frac{E-V_0}{\overline{u}k^\prime}}
{\cos\phi\cos\theta\cos(q_xD)+i\left(\sin\phi\sin\theta-s\frac{E-V_0}{\overline{u}k^\prime}\right)\sin(q_xD)}
\cdot\frac{e^{-i\theta}+se^{i\phi}\frac{E-V_0+\Delta}{\overline{u}k^\prime}}
{e^{-i\theta}-se^{-i\phi}\frac{E-V_0+\Delta}{\overline{u}k^\prime}},
\end{equation}
where $k^\prime=\sqrt{q^2_x+k^2_y}$. The passage to limit,
$\Delta\rightarrow0$, in \eqref{21} is performed by replacements
$\frac{E-V_0}{\overline{u}k^\prime}\rightarrow s^\prime,\
\frac{E-V_0+\Delta}{\overline{u}k^\prime}\rightarrow s^\prime,\
s^\prime=sign(E-V_0)$, the result coincides with formula (4) of Ref.
\cite{Katsnelson}.

The reflection probabilities $R=|r|^2$ for both types of solution in
range II are
\newpage
\begin{equation*}
R_\text{oscil}=\frac{\left[\cos(\phi-\theta)-s\frac{E-V_0}{\overline{u}k^\prime}\right]^2}
{\cos^2\phi\cos^2\theta
\cot^2(q_xD)+\left(\sin\phi\sin\theta-s\frac{E-V_0}{\overline{u}k^\prime}\right)^2}\times
\end{equation*}
\begin{equation}\label{22}
\times\frac{1+2s\frac{E-V_0+\Delta}{\overline{u}k^\prime}\cos(\phi+\theta)+\frac{(E-V_0+\Delta)^2}{\overline{u}^2k^{\prime2}}}
{1-2s\frac{E-V_0+\Delta}{\overline{u}k^\prime}\cos(\phi-\theta)+\frac{(E-V_0+\Delta)^2}{\overline{u}^2k^{\prime2}}},
\end{equation}
\begin{equation}\label{23}
R_\text{damp}=\frac{\overline{u}^2\widetilde{q}^2_x\cos^2\phi+\left(\overline{u}k_y\sin\phi-s(E-V_0)\right)^2}
{\overline{u}^2\widetilde{q}^2_x\cos^2\phi\
\coth^2(\widetilde{q}_xD)+\left(\overline{u}k_y\sin\phi-s(E-V_0)\right)^2}.
\end{equation}
One can see from \eqref{23} that $R_\text{damp}\rightarrow1$ at
$|\widetilde{q}_x|D\gg1$. It is simply verified that the following
equalities are valid
\begin{equation}\label{24}
\begin{split}
T_\text{oscil}+R_\text{oscil}=1,\\
T_\text{damp}+R_\text{damp}=1.
\end{split}
\end{equation}

\vspace{1cm}
\begin{center}
\hypertarget{fig2}{}
\includegraphics[width=4cm]{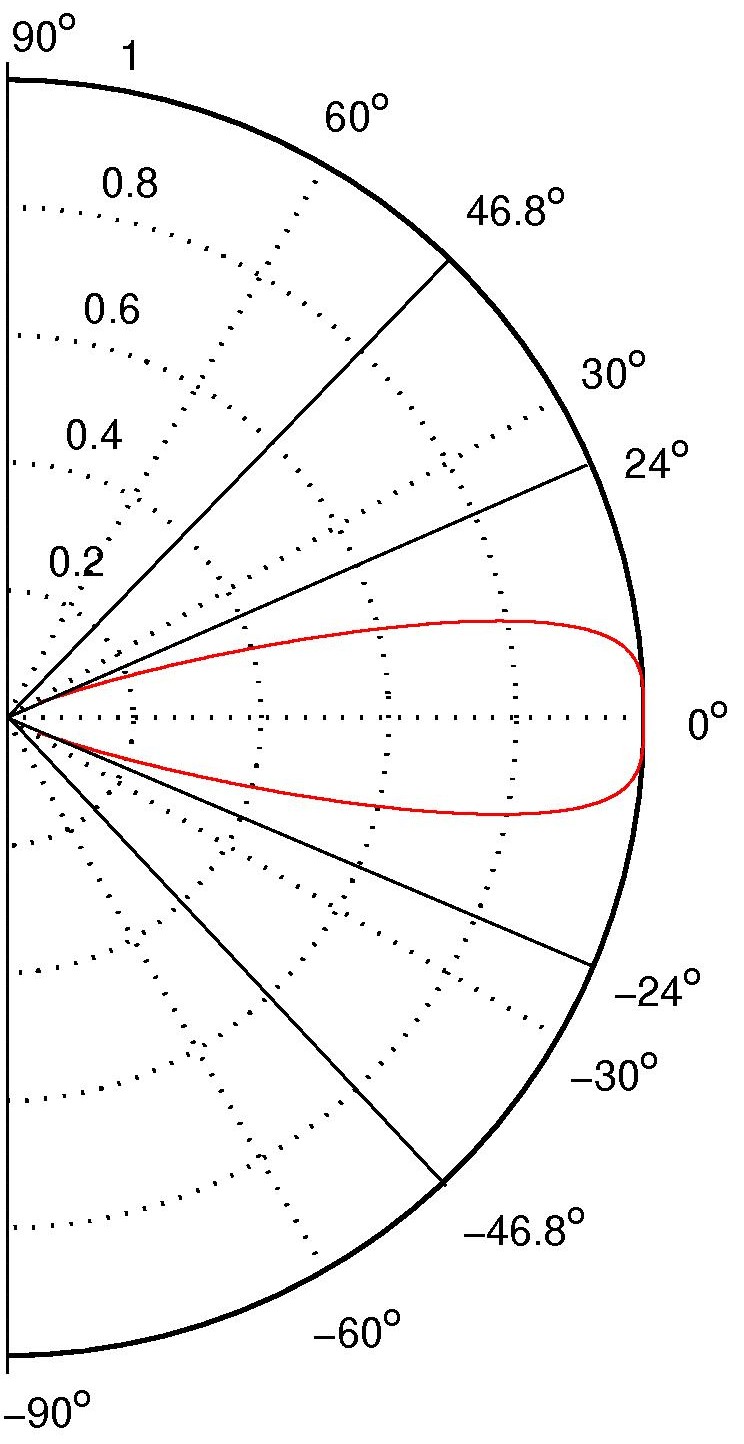}\hspace{-1.5cm}({\bf
a})\hspace{1cm}\includegraphics[width=4cm]{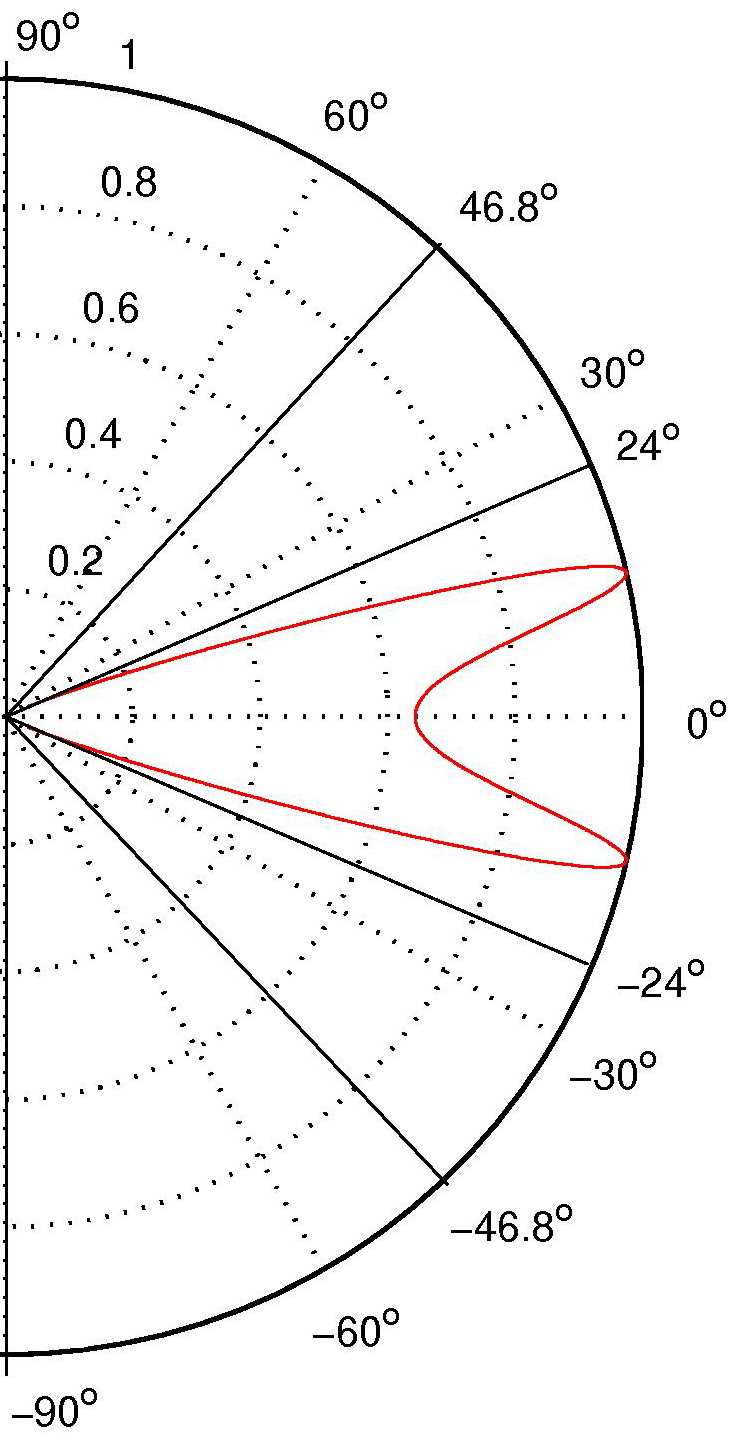}\hspace{-1.5cm}({\bf
b})
\hspace{1cm}\includegraphics[width=4cm]{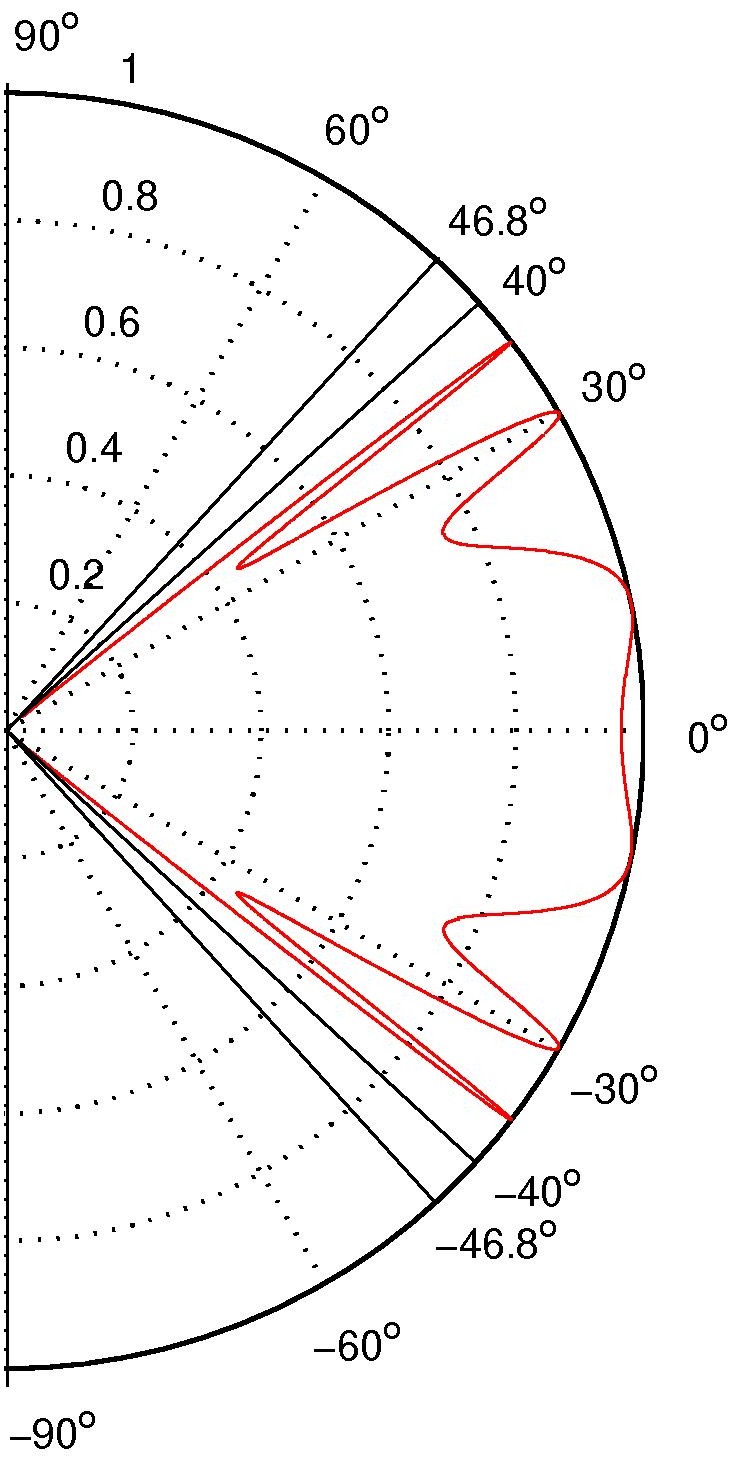}\hspace{-1.5cm}({\bf
c})
\hspace{1cm}\includegraphics[width=4cm]{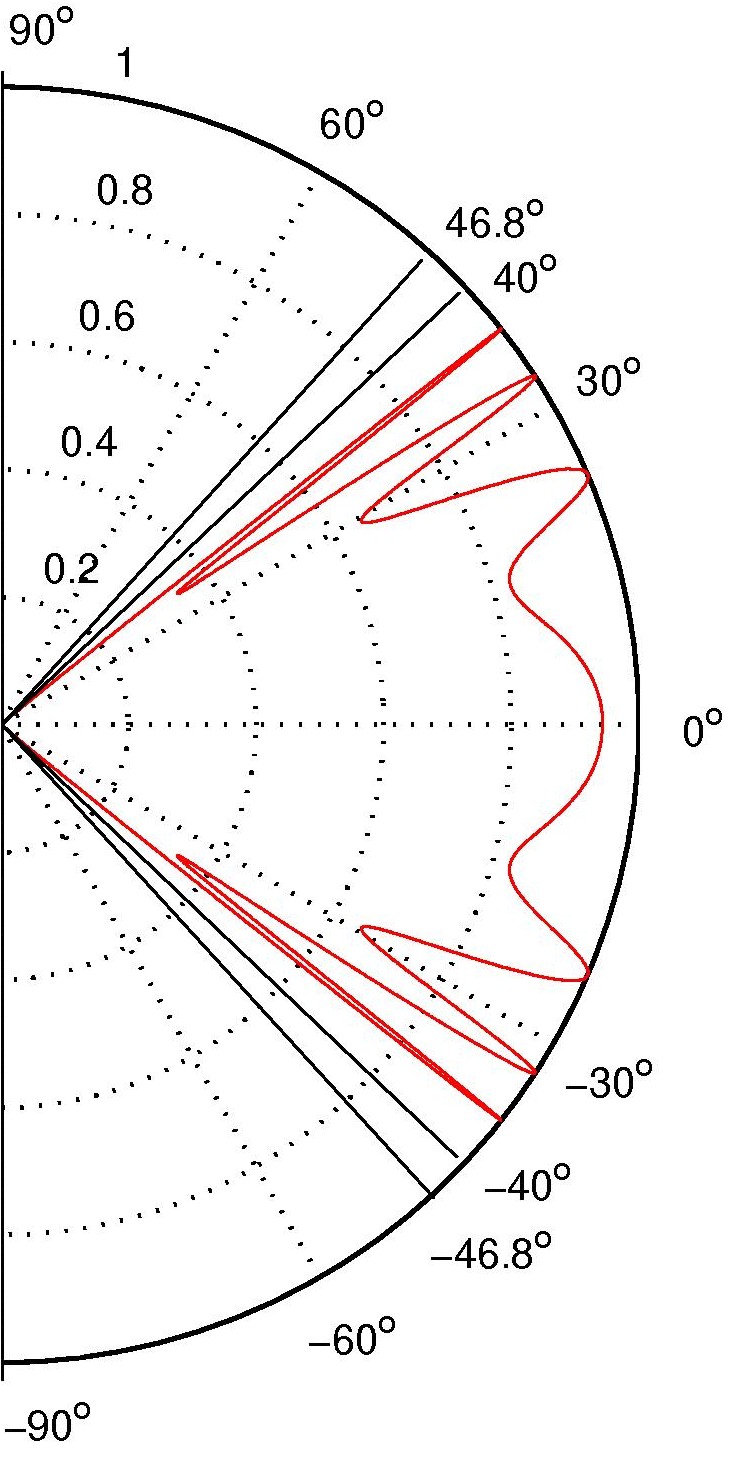}\hspace{-1.5cm}({\bf
d})
\end{center}
\vspace{1cm}Fig. 2. \emph{The dependence of the probability
$T_\text{oscil}$ of the electron transmission through rectangular
barrier being the band gap of the narrow-gap semiconductor GaAs with
$\Delta=705$ meV on the angle of incidence,
$\overline{u}=\sqrt{\frac{\Delta}{m^*}}=1.35\times10^8$ cm/s where
$m^*=0.068m_0, m_0$ is the free electron mass \cite{Silin2}. The
difference of work functions of GaAs and graphene is assumed to be
positive and equal in $V_0=100$ meV. The angle $\phi_0\approx46.8^o$
corresponding to the equality $\sin\phi_0=u/\overline{u}$ is marked.
Two values of energy satisfying the above-barrier transmission
condition $E>\Delta+V_0$ are considered. When the angle of incidence
approaches $\phi_1$, the upper boundary of the above-barrier damped
range comes up to the energy $E$ of a incident electron, for $E=1$
eV $\phi_1\approx24^o$, for $E=2$ eV $\phi_1\approx40^o$: (a) $E=1$
eV, $D=50\ \text{\AA}$; (b) $E=1$ eV, $D=60\ \text{\AA}$; (c) $E=2$
eV, $D=50\ \text{\AA}$; (d) $E=2$ eV, $D=60\ \text{\AA}$.}

\newpage
Let us analyze the conditions at which the oscillating or damped
solution can exist in range II. For definiteness, we consider the
case of electrons\footnote{The case of holes is equivalent to the
case of electrons with an accuracy of the replacement
$E\rightarrow-E$ and $V_0\rightarrow-V_0$.}: its energy $E=uk$ is
positive in graphene, then, for the oscillating solution, the
following equality should hold:
\begin{equation}
uk=V_0+\sqrt{\Delta^2+\overline{u}^2q^2_x+\overline{u}^2k^2_y},
\end{equation}
which is valid at condition
\begin{equation}
uk-V_0>\sqrt{\Delta^2+\overline{u}^2k^2_y}.
\end{equation}
Conversely, it is necessary for the damped solution\footnote{The
expression with sign minus before the square root can also
correspond to the damped solution if $V_0>0$ and value of the square
root is smaller $V_0$.} that
\begin{equation}
uk=V_0+\sqrt{\Delta^2-\overline{u}^2\widetilde{q}^2_x+\overline{u}^2k^2_y}.
\end{equation}
It is valid at condition of the intersection of the dispersion
curves for graphene and the narrow-gap semiconductor
\cite{Kolesnikov*}
\begin{equation}\label{xxx}
uk-V_0<\sqrt{\Delta^2+\overline{u}^2k^2_y}.
\end{equation}
It is evident from the inequality \eqref{xxx} that if the Dirac
point of the BZ of graphene falls into the band gap of the
narrow-gap semiconductor (tunneling through the potential barrier
being the band gap of the narrow-gap semiconductor) then the
solution in range II for electrons with the energy $E_e<V_0+\Delta$
(analogously for holes with the energy $E_h>V_0-\Delta$) is always
damped one.

The momentum range corresponding to the oscillating solution is
defined by inequality
\begin{equation}\label{29}
\left(u^2-\overline{u}^2\sin^2\phi\right)k^2-2V_0uk+V^2_0-\Delta^2>0,
\end{equation}
and the inverse inequality defines the momentum range of the damped
solution. The analysis of inequality \eqref{29} shows the following:

1) if $u>\overline{u}$ then at any angle of incidence
$-\frac{\pi}{2}<\phi<\frac{\pi}{2}$ for electrons with energy $E_e$
and holes with energy $E_h$ in range
\begin{equation}\label{30}
\begin{split}
\Delta+V_0<E_e<E^+(\phi),\\
E^-(\phi)<E_h<-\Delta+V_0,
\end{split}
\end{equation}
where
$E^\pm(\phi)=\frac{V_0\pm\sqrt{\Delta^2-\eta\sin^2\phi(\Delta^2-V^2_0)}}{1-\eta\sin^2\phi},\
\eta=\frac{\overline{u}^2}{u^2},$ there is the above-barrier damped
solution; in range $E_e>E^+(\phi)$ and $E_h<E^-(\phi)$ there exists
the oscillating solution;

2) if $u<\overline{u}$ (it is valid for a number of the narrow-gap
semiconductors, e.g., GaAs and InSb) then it is necessary to
distinguish the following particular cases:

\hspace{0.5cm}  a) the situation in the range of angles
$|\sin\phi|<\frac{u}{\overline{u}}$ is the same as the case 1);

\hspace{0.5cm}  b) the behavior of particles in the range of angles
$\frac{u}{\overline{u}}<|\sin\phi|<1$ is various depending on the
value $V_0$:

\hspace{0.5cm} b*) if
$\Delta\sqrt{1-\frac{u^2}{\overline{u}^2}}<|V_0|<\Delta$ then it
should distinguish the subcases for all values of angles from this
range:

\hspace{1cm} (i) there is the damped solution at any $k$ (at any
energy) for electrons at $V_0>0$ and for holes at $V_0<0$;

\hspace{1cm} (ii) there is energy range above the barrier for
electrons at $V_0<0$ and holes at $V_0>0$, transparency window, in
which there is the oscillating solution, and beyond it there is the
damped solution:
\begin{equation}\label{31}
\begin{split}
E_1(\phi)<E_e<E_2(\phi),\\
-E_2(\phi)<E_h<-E_1(\phi),
\end{split}
\end{equation}
where
$E_{1,2}(\phi)=\frac{V_0\mp\sqrt{\Delta^2-\eta\sin^2\phi(\Delta^2-V^2_0)}}{\eta\sin^2\phi-1}$;

\hspace{0.5cm} b**) if
$|V_0|<\Delta\sqrt{1-\frac{u^2}{\overline{u}^2}}$ then

\hspace{1cm} (j) the situation is the same as case b*) in the range
of angles
$\frac{u}{\overline{u}}<|\sin\phi|<\frac{u}{\overline{u}}\frac{\Delta}{\sqrt{\Delta^2-V^2_0}}$;

\hspace{1cm} (jj) there exists only the damped solution at any $k$
in range of angles
$\frac{u}{\overline{u}}\frac{\Delta}{\sqrt{\Delta^2-V^2_0}}<|\sin\phi|<1$.

The potential barrier is an ideal reflector at sufficiently large
angles of incidence in cases (i) and (jj), i.e. an ``angle filter''
transmitting particles with angles of incidence near to $\phi=0$. At
the same time, it is supposed that $|q_x|D\gg1$, i.e.
$T_\text{damp}\ll1$. Such an unusual feature of the rectangular
potential barrier is related to the circumstance that the ``speed of
light'', analogues of which are $u$ and $\overline{u}$, is different
in graphene and the narrow-gap semiconductor \cite{Kolesnikov}.

The case $u=\overline{u}$ should be attributed to the case 1). Then
the energy range of the above-barrier damped solution disappears,
and a particle behaves as an usual nonrelativistic particle, namely,
there are the damped and oscillating solutions under and above the
barrier, respectively. Similar results in this particular case have
been obtained by Gomes and Peres \cite{Peres}.

Let us consider separately the case when we have instead of the
narrow-gap semiconductor, a gapless semiconductor for which
$\overline{u}\neq u$, and $V_0\neq0$ (at $\overline{u}=u$ this case
coincides with one considered in \cite{Katsnelson}). However, in
contrast to Ref. \cite{Katsnelson}, there is a number of features
distinguished the case $\overline{u}\neq u$. The transmission
probabilities for both types of solution in the gapless
semiconductor are given by expressions \eqref{18} and \eqref{19},
 the only difference is that it is necessary to make the replacement
$E-V_0\rightarrow s\overline{u}k^\prime$. In the above manner, let
us analyze what kind of solution we have in the gapless
semiconductor:

1) if $u>\overline{u}$ then at any angle
$-\frac{\pi}{2}<\phi<\frac{\pi}{2}$

\hspace{0.5cm} a) there exists the oscillating solution for
electrons at $V_0<0$ and for holes at $V_0>0$ for any $k$;

\hspace{0.5cm} b) there exists the damped solution for electrons at
$V_0>0$ and for holes at $V_0<0$ in the energy intervals
\begin{equation}\label{32}
\begin{split}
E^+_0(\phi)<E_e<E^-_0(\phi),\\
E^-_0(\phi)<E_h<E^+_0(\phi),
\end{split}
\end{equation}
where $E^\pm_0(\phi)=\frac{u}{u\pm\overline{u}|\sin\phi|}V_0$; and
there exists the oscillating solution beyond these intervals. If we
regard $V_0$ as the potential barrier height \cite{Katsnelson}, then
we have the under-barrier oscillating solution, this fact
corresponds to the Klein paradox;

2) if $u<\overline{u}$ then

\hspace{0.5cm} a) the situation is the same as in 1) for angles
$|\sin\phi|<\frac{u}{\overline{u}}$;

\hspace{0.5cm} b) for angles $\frac{u}{\overline{u}}<|\sin\phi|<1$,
we are should distinguish two particular cases:

\hspace{1cm} (i) the solution is damped for electrons at $V_0<0$ and
holes at $V_0>0$ for any $k$;

\hspace{1cm} (ii) the solution is oscillating for electrons at
$V_0>0$ and holes at $V_0<0$ in energy ranges \eqref{32} but out of
ones there is the damped solution.

Finally, let us consider the particular case $\Delta=0$ and $V_0=0$
at $\overline{u}\neq u$:

1) if $u>\overline{u}$ then the solution is oscillating at any angle
$-\frac{\pi}{2}<\phi<\frac{\pi}{2}$ and any energy, this fact
corresponds to the Klein paradox;

2) if $u<\overline{u}$ then the solution is oscillating at any
energy for $|\sin\phi|<\frac{u}{\overline{u}}$ and the solution is
damped at $|\sin\phi|>\frac{u}{\overline{u}}$.

In conclusion, we note that the considered heterostructure can be
used as the ``switch'', namely, applying a voltage on the narrow-gap
semiconductor ribbon we can ``switch on'' and ``switch off''
transmission of the charge carriers through range II depending on
the energy range in which the Dirac point of graphene falls (in the
range of the oscillating or damped solution). When we apply an
electric field $F$, the Dirac point of graphene shifts in energy by
the value $\sim eFd$ where $d$ is a distance from the voltage
applying point to the narrow-gap semiconductor ribbon. We suppose
that the electric field is weak enough: $eFd<\Delta-|V_0|$, i.e.
current does not flow at the given $V_0$. The electric field
correction results in displacement $\sim\frac{1}{2}eFD$ of extrema
of the conduction and valence bands of the narrow-gap semiconductor
\cite{Ratnikov3}. Applying the voltage $-U_0$ to the narrow-gap
semiconductor ribbon changes the difference $V^\prime_0=V_0-U_0$ of
work functions between the narrow-gap semiconductor and graphene so
that passage of electrons becomes possible at
$eFd>E^+(\phi)|_{V^\prime_0}$. Condition of passage for holes is
$eFd>|E^-(\phi)|$. Changing $U_0$, we can achieve passage of either
electrons or holes.

An alternative scheme of the ``switch'' is possible. Due to zero gap
in graphene one can pump electrons from the substrate in the
conduction band or displace electrons from graphene thereby
obtaining holes in the valence band. Changing position of the Fermi
level $E_F$ in one of the graphene layers we can provide passage of
either electrons at the condition $eFd+E_F>E^+(\phi)$ ($E_F>0$) or
holes at $-eFd+E_F<E^-(\phi)$ ($E_F<0$).

\flushright Received 29 May 2008.
\end{document}